\documentclass{article} 

\usepackage[hidelinks]{hyperref}
\pdfoutput=1
\usepackage{arxiv}
\usepackage[utf8]{inputenc} 
\usepackage[T1]{fontenc}    
\usepackage{amsfonts}       
\usepackage[intlimits]{amsmath}
\usepackage{nicefrac}       
\usepackage{microtype}      
\usepackage{doi}

\usepackage{chemformula} 

\usepackage{siunitx}

\usepackage{chngcntr}

\usepackage{lineno}
\usepackage{float}
\usepackage{cleveref}

\usepackage{tikz}

\title{Additive GaN solid immersion lenses for enhanced photon extraction efficiency from diamond color centers}

\author{Xingrui Cheng$^*$\\
    \textbf{Andrew R. Kirkpatrick}\\
    \textbf{Jason M. Smith}\\
    \textbf{Patrick S. Salter}\\
	Department of Engineering Science\\
    and Department of Materials\\
    University of Oxford\\
    Oxford, UK\\
	\And
    Nils Kolja Wessling$^*$\\
    \textbf{Martin D.\,Dawson}\\
    \textbf{Michael J. Strain}\\
	Institute of Photonics\\
	Department of Physics\\
	University of Strathclyde \\
    Glasgow, UK\\
    \textit{michael.strain@strath.ac.uk}
    \And
	Saptarsi Ghosh\\
    \textbf{Menno Kappers}\\
    \textbf{Rachel\,A.\,Oliver}\\
	Cambridge Centre for Gallium Nitride\\
    University of Cambridge\\
    Cambridge, UK
    \And
    Yashna N. D. Lekhai\\
    \textbf{Gavin W. Morley}\\
    Department of Physics\\
    University of Warwick\\
    Coventry, UK\\
}

\date{}

\renewcommand{\headeright}{}
\renewcommand{\undertitle}{}
\renewcommand{\shorttitle}{Additive GaN solid immersion lenses for enhanced photon extraction efficiency from diamond color centers}

\hypersetup{
pdftitle={Additive GaN solid immersion lenses for enhanced photon extraction efficiency from diamond color centers},
pdfsubject={physics.app-ph},
pdfauthor={Xingrui Cheng, Nils Kolja Wessling, Saptarsi Ghosh, Andrew R.
Kirkpatrick, Menno J. Kappers, Yashna N. D. Lekhai, Gavin W. Morley, Rachel A. Oliver, Jason M. Smith, Martin D. Dawson, Patrick S. Salter, Michael J. Strain},
pdfkeywords={Diamond, nitrogen vacancy centre, additive GaN micro-optics, transfer printing, quantum system engineering}
}

\begin{document}

\maketitle

\begin{abstract}
	Effective light extraction from optically active solid-state spin centres inside high-index semiconductor host crystals is an important factor in integrating these pseudo-atomic centres in wider quantum systems. Here we report increased fluorescent light collection efficiency from laser-written nitrogen vacancy centers\,(NV) in bulk diamond facilitated by micro-transfer printed GaN solid immersion lenses. Both laser-writing of NV centres and transfer printing of micro-lens structures are compatible with high spatial resolution, enabling deterministic fabrication routes towards future scalable systems development. The micro-lenses are integrated in a non-invasive manner, as they are added on top of the unstructured diamond surface and bond by Van-der-Waals forces. For emitters at 5\,$\mu$m depth, we find approximately 2$\times$  improvement of fluorescent light collection using an air objective with a numerical aperture of NA\,$=0.95$ in good agreement with simulations. Similarly, the solid immersion lenses strongly enhance light collection when using an objective with NA\,$=0.5$, significantly improving the signal-to-noise ratio of the NV center emission while maintaining the NV's quantum properties after integration.
\end{abstract}

\begin{minipage}{25em}
\vspace{0.5cm}
*Both authors contributed equally to this work
\end{minipage}


\section{Introduction}
\label{sec:Intro}
Solid state quantum defects in bulk crystals have emerged as promising systems for applications such as quantum information science\,\cite{Wolfowicz2021,Weber2010,childress2013diamond}, quantum sensing\,\cite{Lounis2005,Barry2020} and quantum imaging\,\cite{Scholten2021}. These defects are particularly attractive for such applications due to their long spin coherence times, room temperature stability, and the possibility for scalable fabrication and integration into existing electronic and photonic technologies.\\
Among these solid-state quantum defects, the negatively charged nitrogen vacancy center (NV$^-$) in diamond has gained particular attention due to beneficial optical and spin properties. NV$^-$ centers are point defects in the diamond lattice that consist of a substitutional nitrogen atom adjacent to a carbon vacancy site. They exhibit bright and stable photoluminescence\,(PL) at room temperature, as well as long-lived electron spin states which can be used as auxiliary qubits to nearby nuclear spins\,\cite{abobeih2022fault}, nanoscale magnetic field sensors\,\cite{maletinsky2012robust} or nodes in a quantum network\,\cite{johnson2017diamond,pompili2021realization}.\\
In recent years there has been significant progress in developing techniques for fabricating NV centers in diamond, including ion implantation\,\cite{Cui2012,Sangtawesin2014},  and low energy electron beam irradiation\,\cite{Chu2014,McLellan2016} with post-treatment annealing. Laser writing has emerged as a particularly attractive fabrication technique due to the unique non-linear light-matter interaction mechanism so that single NV$^-$ centers can be fabricated deterministically with minimal residual lattice damage and precision positioning inside host materials\,\cite{Chen2019,Chen2017,hadden2018integrated,Wang2022}.\\
The collection efficiency of photons emitted by diamond color centers is a major limitation for many quantum applications, affecting other interesting emitters such as the silicon\,(SiV), tin\,(SnV) or germanium vacancy\,(GeV) center similarly\,\cite{bradac2019quantum,debroux2021quantum,chen2022quantum,malykhin2022control,ju2021preparations,janitz2020cavity}. The high refractive index mismatch between diamond and the surrounding medium results in a considerable proportion of emitted photons being reflected internally, leading to a significant loss of emission signal from any emitter inside the crystal. In order to address this limitation, several techniques have been developed to efficiently address atomic defects inside single crystalline diamond, including cavity coupling\,\cite{Barclay2009,johnson2015tunable,gould2016efficient,Dolan2018} and nanostructuring\,\cite{babinec2010diamond,castelletto2011diamond,li2015efficient,sipahigil2016integrated,radtke2019nanoscale,jeon2020bright}.\\
A particularly common approach to increase the light collection efficiency is the fabrication of a solid immersion lens (SIL) around a defect, typically by using a focused ion beam\,(FIB) to carve the diamond into a lens shape\,\cite{Hadden2010,marseglia2011nanofabricated, Jamali2014,Rogers2014}. A standard milling process is stated to take around 1\,h per lens\,\cite{Jamali2014} and is therefore challenging to scale, but significant collection improvements of up to 10$\times$ have been reported. Even though no detrimental effects of the milling process on the coherence properties of NV$^-$ centers are reported\,\cite{jiang2014focused}, there are some concerns that the ion bombardment might cause additional strain around the defect of interest\,\cite{Wang2020,Knauer2020}. Recent work at the forefront of NV-based quantum computing and quantum networking reports the use of such monolithic SILs for faster measurement time and reduced error rate due to enhanced signal to noise ratio, illustrating the practical significance of this method\,\cite{abobeih2022fault,pompili2021realization}.\\
In this paper the advantages of laser written NV$^-$ centers are combined with back-end integration of GaN solid immersion lenses using an additive micro-assembly method, avoiding any damage to the host crystal. This heterogeneous integration approach allows the decoupling of the lens fabrication processing from the vacancy centre definition and selection, while massively speeding up the lens fabrication due to the use of wafer scale compatible parallel wet and dry etching techniques. The GaN lenses are realised using the balanced etch selectivity of photoresist and GaN in inductively coupled plasma\,(ICP) etching creating high aspect ratio lenses, which cannot commonly be achieved when processing diamond with ICP etching\,\cite{Wessling2022}. GaN and diamond are closely index matched around the emission wavelength of the NV$^-$ center providing minimal reflection effects at the material interface. The detailed fabrication process is outlined in the following section. 
\section{Fabrication and lens integration}
\begin{figure}[H]
    \centering
    \includegraphics[width=\linewidth]{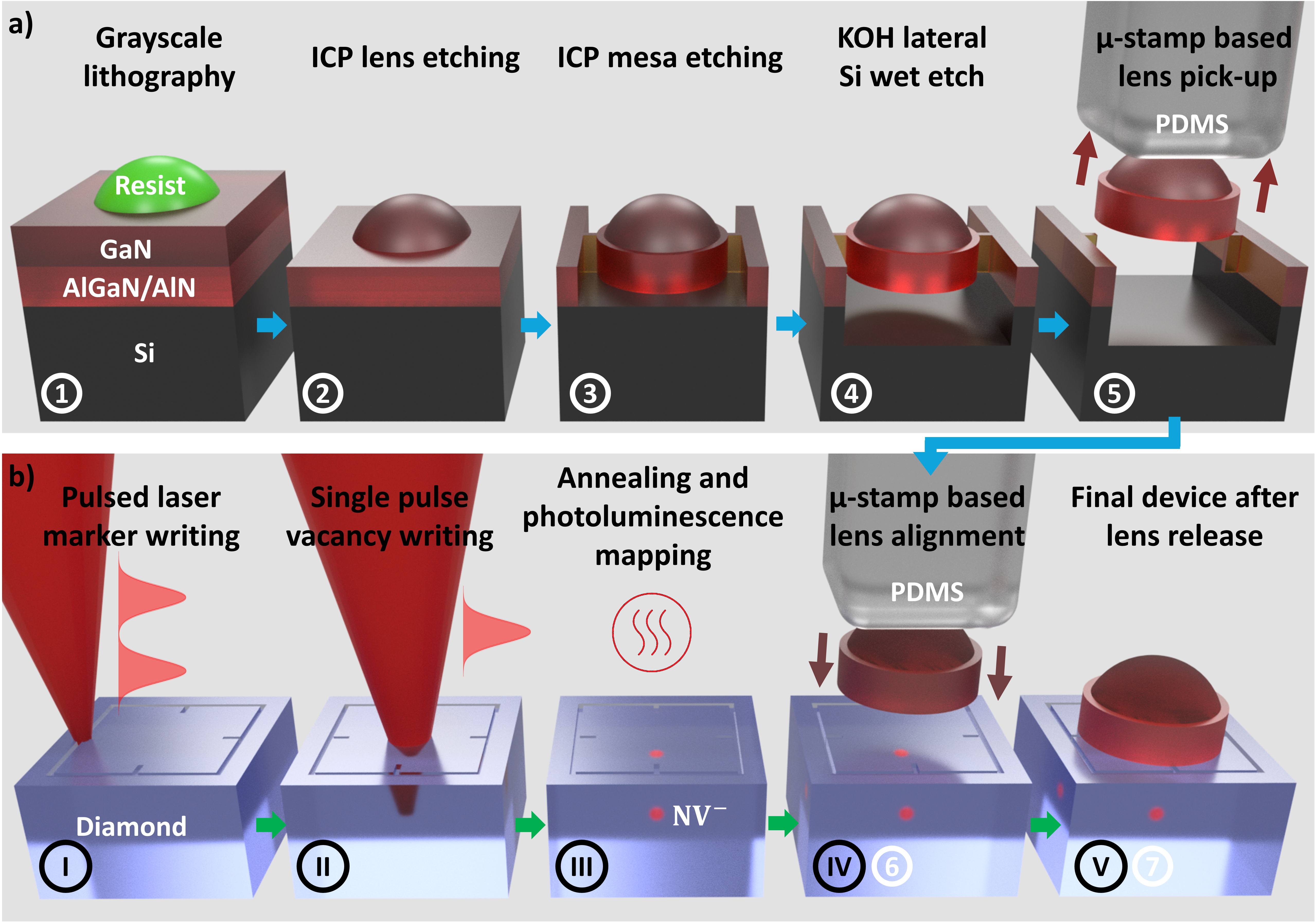}
    
\caption{Schematics describing the process flow, (a) suspended GaN solid immersion lenses are fabricated by parallel plasma dry and wet etching techniques, (b) after initial marker writing, a NV$^-$ defect center is generated in single crystalline diamond by pulsed laser writing and subsequent thermal annealing, and the suspended GaN lenses are assembled above the defect center with micro-transfer printing.}
\label{fig:Process}
\end{figure}

\begin{figure}[H]
\centering
\includegraphics[width=\linewidth]{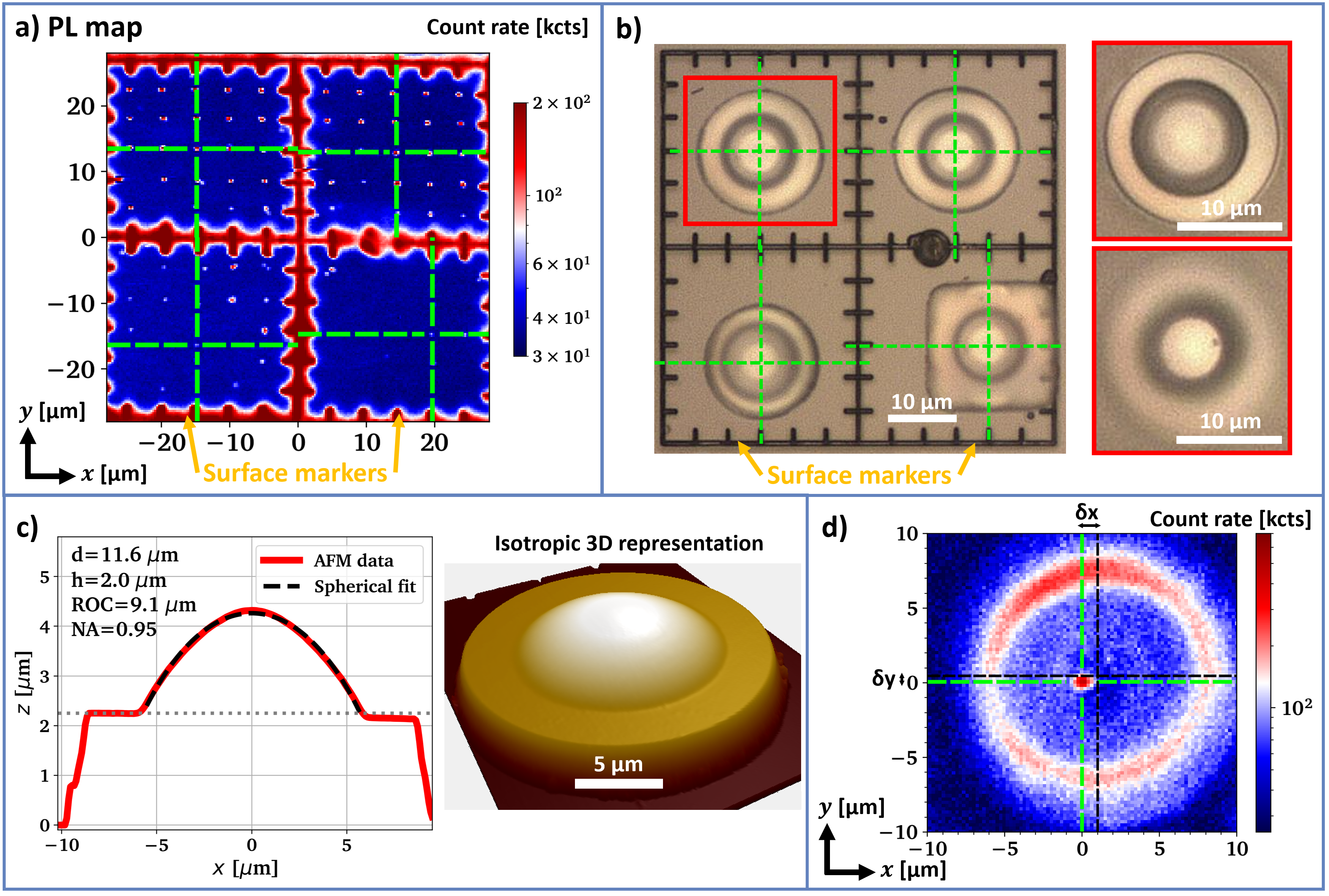}
\caption{a) Photolumincescence image of an example region after laser writing and annealing. The image was recorded using an oil immersion objective with NA\,$=1.25$. Targeted emitter locations are marked with green crosses, b) Microscope images of the assembled GaN SILs above the marked emitters in a), c) AFM data of the GaN SIL with the red frame (top left quadrant) in b), d) Photolumincescence image of a graphitised spot below the SIL from c), recorded using an air objective with NA\,$=0.5$.}
\label{fig:Fabrication}
\end{figure}

A schematic of the process flow is indicated in Fig.\,\ref{fig:Process} while corresponding experimental results are shown in Fig.\,\ref{fig:Fabrication}. 
Fig.\,\ref{fig:Process}\,a) illustrates how the GaN solid immersion lenses are defined and suspended on a strain-optimised heteroepitaxially grown GaN-on-Si chip\,\cite{spiridon2021method,ghosh2021origin,ghosh2023design} based on the wafer-scale compatible microfabrication process reported in earlier work\,\cite{Wessling2022}. Initially polymer resist lenses are defined by grayscale lithography and thermal reflow\,(1), followed by inductively coupled plasma\,(ICP) dry etching to transfer the lens shape into the 2\,$\mu$m thick GaN top layer\,(2). A mesa structure is lithographically defined around the lens and translated into the ca. 2\,$\mu$m thick AlGaN/AlN buffer layer in an additional ICP etching step, exposing the Si substrate\,(3). Utilising a SiO$_x$ hard mask, the lens and AlGaN/AlN mesa are laterally suspended by an anisotropic potassium hydroxide\,(KOH) wet etch followed by hard mask removal\,(4). The prefabricated GaN SILs are then extracted from their growth substrate using a modified dip-pen lithography system\,\cite{mcphillimy2018high} under white light illumination\,(5). A patterned polydimethylsiloxane micro-stamp (6:1 Sylgard 184 PDMS) with around 35\,$\mu$m extrusion height and 30\,$\mu$m square lateral size is used. This polymer-based micro-transfer printing technique allows the deterministic release and placement of suspended semiconductor chiplets with sub-micron lateral precision using the reversible adhesion properties of PDMS\,\cite{feng2007competing, mcphillimy2018high, guilhabert2022advanced}.\\
The fabrication of diamond nitrogen vacancy centers at targeted positions and successive GaN SIL integration is illustrated in Fig.\,\ref{fig:Process}\,b). A commercially available electronic-grade single crystalline [100] diamond grown by chemical vapour deposition with nitrogen density $<5$\,ppb is used as substrate. Laser writing is implemented using a regeneratively amplified Ti:Sapphire source producing laser pulses with 150\,fs duration, 790\,nm wavelength and a repetition rate of 1\,kHz which is focused using a high-NA oil immersion objective lens\,(Olympus 60$\times$, 1.4\,NA). Initially a high laser pulse energy is used to break down the diamond lattice and create square graphite markers in regions with low intrinsic color center concentration\,(I). These surface markers are easily visible in an optical microscope, aiding the localisation of the writing sites and later the transfer printing process, as shown in Fig.\,\ref{fig:Fabrication}\,b).

A single pulse from the femtosecond laser is focused inside the electronic grade diamond substrate to introduce an ensemble of Frenkel defects at the target position determined by the markers in 5\,$\mu$m depth\,(II). Optical aberrations related to refraction at the diamond interface are corrected using a liquid crystal spatial light modulator\,\cite{simmonds2011,griffiths2021}. The sample is then subjected to a \SI{1000}{\degreeCelsius} anneal for 3 hours under nitrogen flow to mobilise the vacancies, some of which combine with intrinsic substitutional nitrogen atoms in the diamond lattice to form stable NV centers\,(III)\,\cite{Chen2017, stephen2019}.\\
The laser-processed areas of the diamond sample are initially characterised by collecting the photoluminescence\,(PL) emission with a home-built confocal microscope using an oil immersion objective\,(NA\,$ = 1.25$)\,(III). A 532\,nm continuous-wave\,(CW) laser\,(GEM 532) is used as excitation source and the collected epi-fluorescence signal is focused on a single photon avalanche diode detector\,(SPAD), filtering in the spectral range from 600\,nm to 740\,nm. The setup is described in more detail in Fig.\,B.1 in the Supplementary Information. Photoluminescence maps of the created emitters were obtained at room temperature, with an example shown in Fig.\,\ref{fig:Fabrication}\,a). As can be seen here, each marker quadrant contains an array of 5$\times$5 writing sites with 5\,$\mu$m spacing. Each site is irradiated by a single laser pulse, with pulse energy kept constant within a particular row but modified between rows. The laser pulse energy range straddles the narrow transition between the regime of lattice breakdown and graphitisation to the regime of creating vacancy ensembles without any sp$^2$-bonded carbon content. The in-plane placement accuracy of the NV centers is expected to be within 250\,nm with respect to the grid\,\cite{Chen2017}. The upper two quadrants of the array typically contain graphitised points while the lower two quadrants show vacancy ensembles. Anti-bunched photon emission is identified by collecting $g^2(\tau)$\,autocorrelation statistics using a second SPAD and the positions of promising emitters are noted to guide the transfer printing of the GaN micro-lenses.

Using the detailed spatially resolved precharacterisation of the photo-emitters, GaN SILs with a radius of curvature\,(ROC) matching the emitter depth are preselected, removed from their growth substrate and deterministically placed above targeted NV$^-$ centers using the laser written marker structures as alignment guide, as illustrated in Fig.\,\ref{fig:Process}\,a)\,(IV)/(6). The related experimental results after lens release\,(V)/(7) are shown in Fig.\,\ref{fig:Fabrication}\,b). Here, multiple micro-lenses are integrated on a small footprint, with a single GaN lens placed in each quadrant of a pre-characterised marker region. The  corresponding PL map shown in Fig.\,\ref{fig:Fabrication}\,a) is used to adjust the print position according to the most promising emitter locations. Alignment is mainly limited by image distortions in the PL map and the markers are solely used for visual guidance without applying any numerical methods. Before transfer printing, an ultrasonic solvent clean and a boiling acid treatment in 95\,\% sulfuric acid mixed with 30\,\% H$_2$O$_2$ with 3:1 volume ratio is applied to the diamond surface to remove any surface debris remaining after laser writing the surface markers. The adhesion of the GaN micro-lens to the diamond surface relies solely on Van-der-Waals forces without the use of any adhesion layers. This type of bonding depends on both $\mu$m-scale flatness and nm-scale local roughness of the participating surfaces. The micro-lens height is restricted to the 2\,$\mu$m thick GaN epilayer to achieve a flat bottom surface\,\cite{Wessling2022}.\\
As stated in earlier work\,\cite{Wessling2022} a combination of grayscale lithography and resist reflow is used to fabricate spherical and smooth lens profiles with highly engineered lens dimensions. Fig.\,\ref{fig:Fabrication}\,c) shows an atomic force microscopy\,(AFM) line scan and a true-scale 3D representation of the AFM data of the marked device in Fig.\,\ref{fig:Fabrication}\,b), indicating a smooth symmetrical lens with a fitted ROC\,$=9\,\mu$m, taking 4\,$\mu$m overall device thickness into account. After lens integration, we reevaluate the photoluminescence emission from the targeted emitter ensembles though the GaN SILs using air objective lenses. Fig.\,\ref{fig:Fabrication}\,d) shows the emission from a graphitised spot in the top left quadrant using an objective lens with NA\,$=0.5$. The SIL is expected to contribute a magnification similar to its refractive index\,$n=2.4$ if the emitter is placed in the geometric centre of the lens sphere\,\cite{baba1999aberrations}. The visible lateral displacement between emitter and lens center accounts to roughly ($\delta x,\delta y)=(1.0,0.4)\,\mu$m, indicating a real displacement of ($\Delta x,\Delta y)=(0.4,0.2)\,\mu$m. Our simulations show, that the gross expected collection enhancement is maintained for both collection optics with NA\,$=0.5$ and NA\,$=0.95$ if the real lateral displacement does not exceed $\pm1\,\mu$m, see Fig.\,B.3 in the Supplementary Information.\\ 
The PL map in Fig.\,\ref{fig:Fabrication}\,d) reveals significant background fluorescence from the edges of the lens but the much weaker light emission from the center of the GaN lens is largely rejected by the confocal microscope arrangement. The photoluminescence emission spectrum of the GaN/AlGaN/AlN layer stack under green CW laser excitation is broad, covering 550-800\,nm wavelength at room temperature, compare Fig.\,B.7 in the Supplementary Information. Therefore spectral filtering can only be partially applied to isolate the NV$^-$ emission. Noticeably the apparent emitter depth increases due to the printed SIL in agreement with expectation: Without lens, refraction at the planar diamond interface causes the apparent emitter depth to lie much closer to the diamond air interface than its actual position inside the crystal\,\cite{Jamali2014}.\\
Overall, we demonstrate three dimensional deterministic matching of emitters and SILs using targeted color center laser writing, grayscale lens-shape control and micro-transfer printing, moving towards novel micro systems development with scaling potential.

\section{Simulated light extraction efficiency}
\begin{figure}[H]
    \centering
    \includegraphics[width=\linewidth]{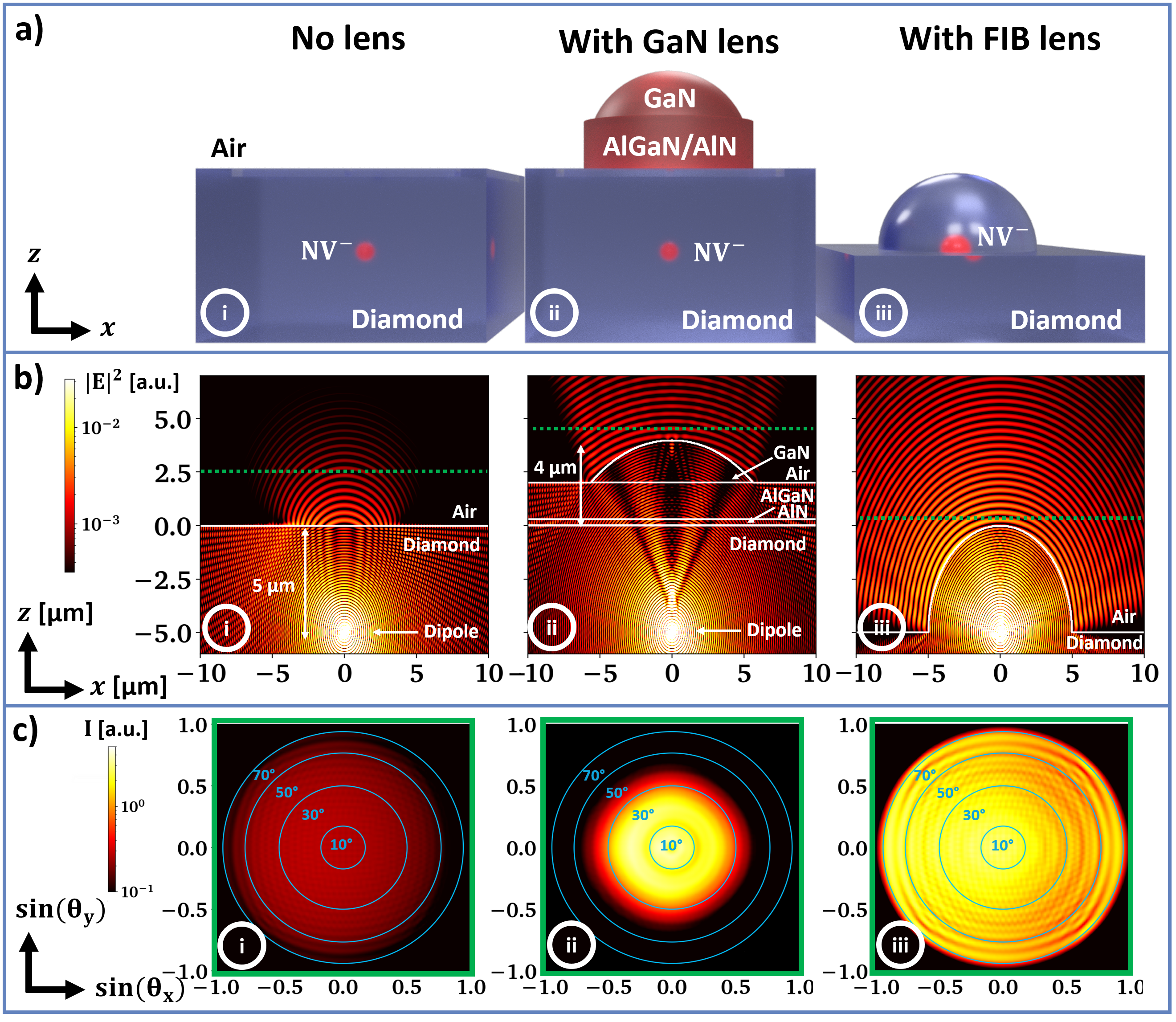}
    
\caption{a) Schematics introducing three different cases of NV-to-free-space coupling with b) corresponding device cross sections from finite difference time domain simulations at $\lambda=700$\,nm wavelength with two dipole emitters mimicking the NV defect center in 5\,$\mu$m depth and c) the far field projection in free-space above the emitter derived from the simulations in b), similarly taken at $\lambda=700$\,nm wavelength.}

\label{fig:Motivation}
\end{figure}
In order to quantify the potential collection efficiency improvement from NV$^-$ centers caused by the transfer printed GaN SILs, three different scenarios are investigated with finite difference time domain simulations\,(FDTD): (i) a flat diamond substrate, (ii) a flat diamond substrate with added GaN micro-lens and (iii) a monolithic hemispherical solid immersion lens fabricated around the emitter, as can typically be achieved by FIB milling. Fig.\,\ref{fig:Motivation}\,a) illustrates all three cases. The GaN lens in simulation (ii) closely resembles the fabricated micro-lens shown in Fig.\,\ref{fig:Fabrication}\,c). The emitter position overlaps with the midpoint of the lens sphere for both the GaN and the diamond lens. The commercial simulation software "Ansys Lumerical FDTD" is used with absorbing boundary conditions.\\
In all three scenarios, the two electric dipole moments of the nitrogen vacancy center are modeled by two dipole emitters perpendicular to the tilted NV$^-$ symmetry axis at 5\,$\mu$m depth below the (100) diamond surface. Fig.\,\ref{fig:Motivation}\,b) shows the intensity cross\,section through each simulation region. In the case of the flat diamond surface the emitted wavefront exhibits strong curvature after passing the diamond-air interface, indicating large angles of the \textbf{$k
$}-vector towards the surface normal caused by refraction. Additionally, total internal reflection\,(TIR) traps significant amounts of the upwards traveling light in the diamond slab. In contrast, both GaN and monolithic SILs allow the wavefront to maintain its shape after passing through the semiconductor materials because their radius of curvature\,(ROC) is matched to the position of the dipole emitter. Additionally, the GaN micro-lens visibly reduces the negative impact of total internal reflection while the FIB lens can eliminate TIR fully.\\
The upwards directed far-field emission pattern is recorded using monitors positioned at the green lines in Fig.\,\ref{fig:Motivation}\,b), assessing the theoretical collection efficiency and its dependence on the numerical aperture\,(NA) of the collection optics. The result of the far field projections is displayed in Fig.\,\ref{fig:Motivation}\,c). As expected the intensity distribution is comparably weak and spread over wide angles if no SIL is in place, while both SILs increase the maximum light intensity by about one order of magnitude. Due to the buffer layer thickness and additive nature of the assembly process, the far field above the GaN micro-lens indicates strong improvement primarily when collection optics with low numerical aperture\,(NA\,$<0.6$) are considered. For collection optics with NA $>0.6$, not much additional gain in absolute collection efficiency is expected. Lower numerical aperture collection optics could be used to address arrays of multiple quantum emitters in a significantly enlarged field of view, which might offer a route to scaling of such quantum systems using spatial light modulators and free-space emitting photonic integrated circuits for beam delivery\,\cite{christen2022integrated}.

\section{Analysis of signal enhancement}
\label{sec:PLenhancemnet}

\begin{figure}[H]
\centering
\includegraphics[width=\linewidth]{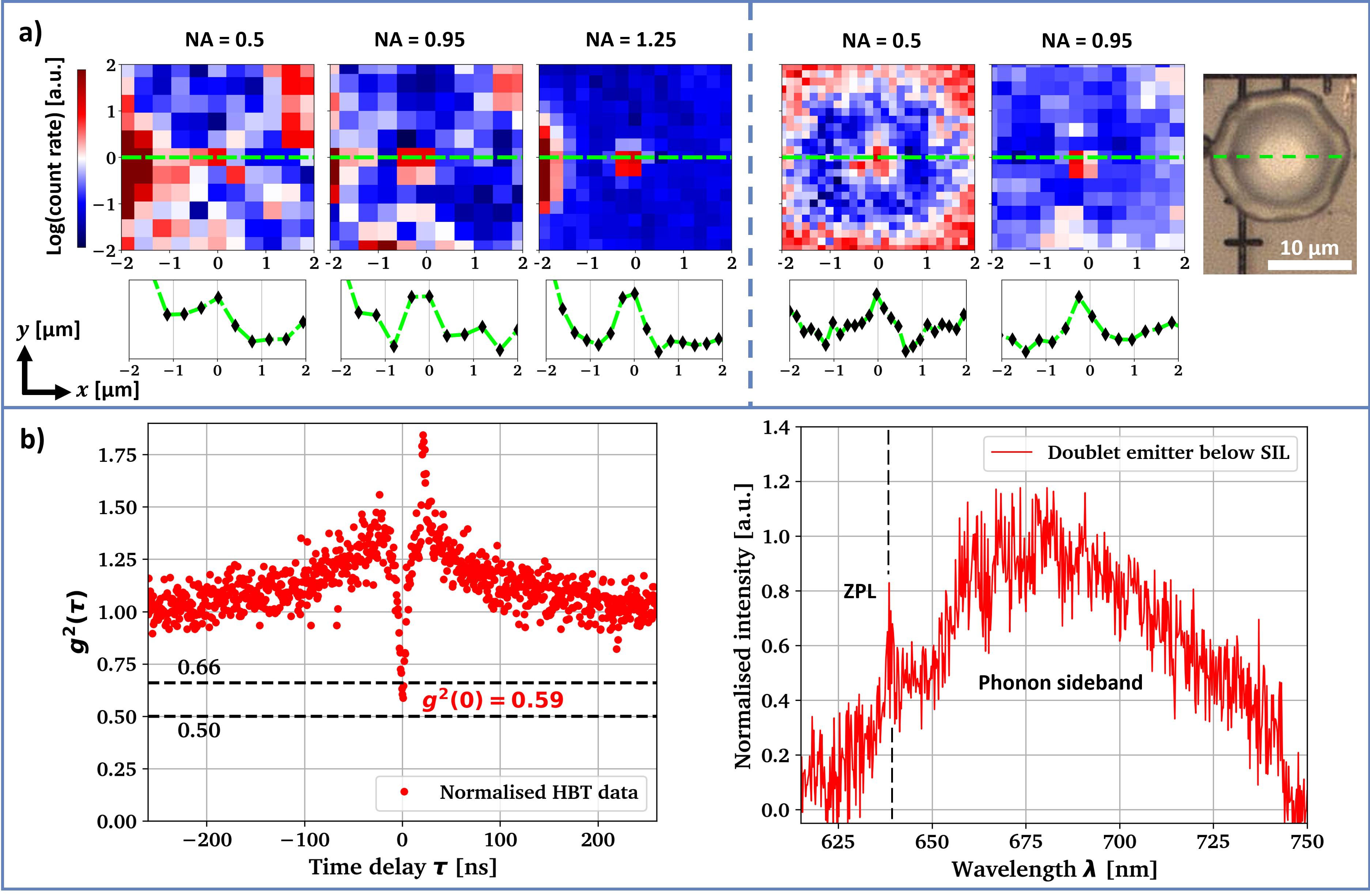}
\caption{a) PL images of the same emitter after the initial laser writing and annealing (left) and after GaN SIL integration (right) showing dependence on the numerical aperture of the collection optics. Line profiles are taken along the green lines in the maps, b)\,normalised autocorrelation without background correction and background corrected spectral data from the emitter discussed in a) after SIL integration using the objective with NA\,$=0.95$.}
\label{fig:Fluorescence}
\end{figure}

The improvement of PL collection efficiency is assessed by comparing two sites where we combined NV$^-$ center pairs with GaN lenses and show that anti-bunching in the photon statistic is maintained after lens integration. Two different home-built confocal microscopes are used to assess the effects of varying NA of the collection optics\,(NA\,$ = 0.5, 0.95$ and 1.25), and both setups are depicted schematically in Fig.\,B.1 and\,B.2 in the Supplementary Information. The setups are qualitatively similar, but we cannot directly compare count rates quantitatively between them.\\
Fig.\,\ref{fig:Fluorescence}\,a) includes PL maps of one writing site which is identified as a NV$^-$ center pair. The PL signal is compared for different objective lens NAs with and without a GaN SIL in place. Prior to SIL integration, the written site is barely visible when imaged with NA\,$=0.5$, but with increasing the objective lens NA, the signal to noise ratio\,(SNR) increases significantly due to more efficient collection from the heavily refracted emitted light. After adding a GaN micro-lens on top of the same emitter, we are now able to clearly resolve the emission using NA\,$=0.5$, noting a roughly 5$\times$ enhanced count rate when the emitter is pumped to saturation. But the low SNR before lens integration makes it difficult to quantify the improvement with much accuracy. The magnification effect of the SIL additionally separates the emitter more clearly from the emitting surface marker structure in the left part of the PL map. The logarithmic line scans through the emitter PL signal reveal significantly improved SNR when comparing both NA\,$=0.5$ and NA\,$=0.95$ before and after lens integration.\\
To confirm the compatibility of the GaN micro-lenses with measurements in the quantum regime, $g^2(\tau)$ autocorrelation measurements are taken after lens integration. Intensity autocorrelation is a well-established technique often used to verify the existence of single photon sources in solid state physics and goes back to the work of Hanbury Brown and Twiss\,(HBT)\,\cite{brown1954lxxiv,Maze2008,Berthel2015}. To collect the $g^2(\tau)$ statistic, the PL signal is split between a free-space SPAD and a fiber-coupled SPAD with timing jitter $\Delta\tau\approx500$\,ps using a 50:50 beamsplitter. For these measurements the objective lens with NA\,$=0.95$ is used. A spectral filter in the optical path of one SPAD prevents potential optical cross-talk due to the breakdown flash commonly observed in Si APDs and SPADs\,\cite{kurtsiefer2001breakdown}.\\
Fig.\,\ref{fig:Fluorescence}\,b) contains the HBT measurement result from the emitter depicted in Fig.\,\ref{fig:Fluorescence}\,a) and we find $0.5<g^2(0)=0.59<0.66$ after lens integration, indicating that the ensemble consists of two closely spaced NV$^-$ centers, which cannot be resolved separately. Fig.\,\ref{fig:Fluorescence}\,b) also includes a background corrected spectral measurement of the discussed emitter taken through the lens, exhibiting  a sharp zero-phonon line (ZPL) centerd at 637\,nm and a phonon sideband of approximately 100\,nm width, as is characteristic of the NV$^-$ center in diamond \,\cite{gruber1997scanning,Jelezko2006}. The long wavelength transmission edge of the 665/150\,nm band-pass filter can be seen around 740\,nm wavelength. For the $g^2(\tau)$ measurement, a 600\,nm long-pass filter is added to remove the first order diamond Raman line from the background signal. Both measurements demonstrate that the photophysics of the emitters remain undisturbed after passing through the GaN SIL. Furthermore, the GaN SIL enhances the photon count rate to enable faster measurements with similar SNR.

\begin{figure}[H]
\centering
\includegraphics[width=\linewidth]{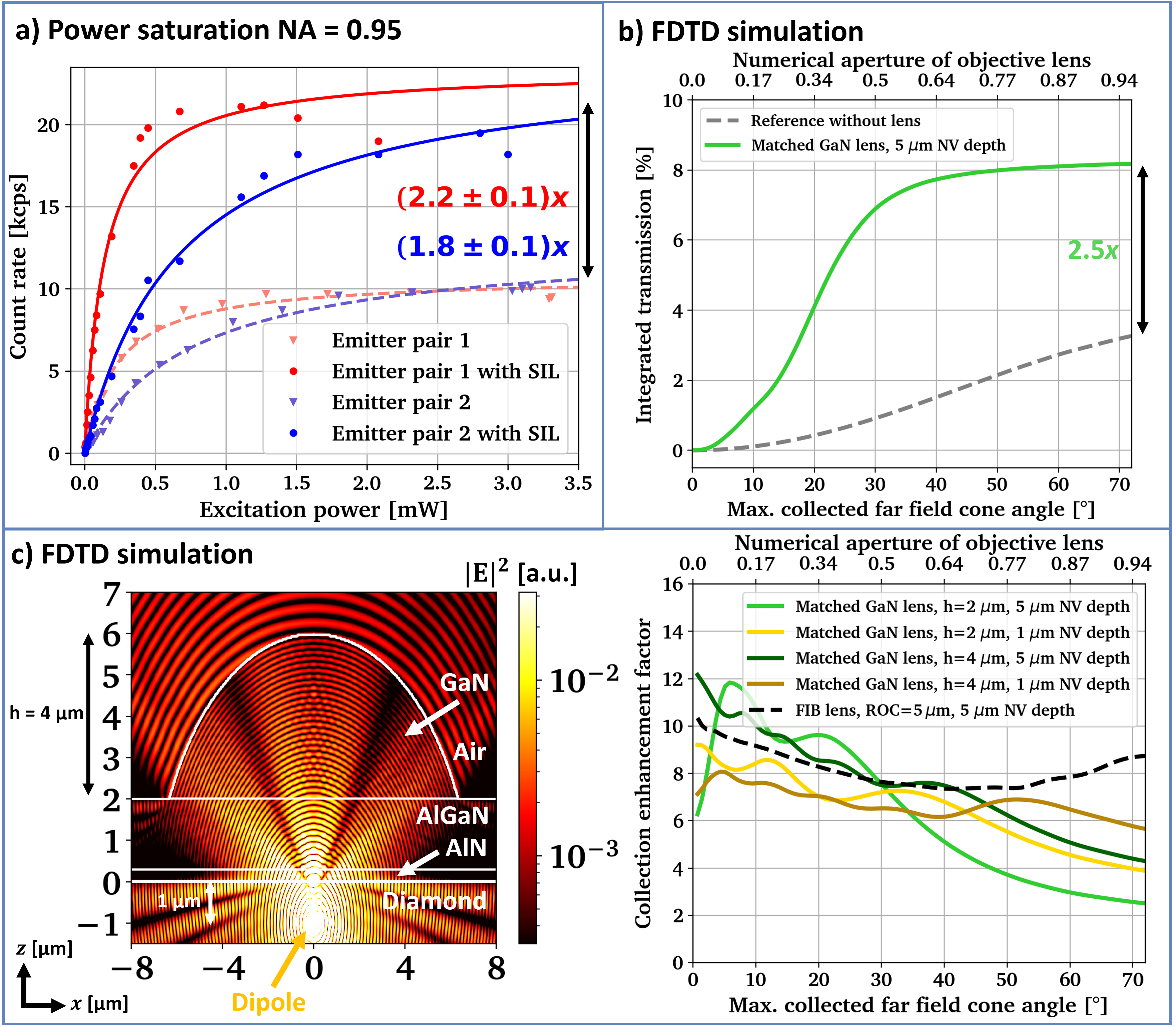}
\caption{a) Fitted power saturation measurements of two NV$^-$ center pairs with and without GaN SIL using an objective with NA\,$=0.95$, b) expected free space collection improvement in dependence of the objective's NA derived from the FDTD simulations shown in Fig.\,2, c) cross section though a simulation region with a GaN SIL matched to a dipole emitter at $\lambda=700$\,nm wavelength and the expected corresponding free space collection enhancement in reference to a planar diamond-air interface in dependence of the emitter depth and GaN lens height. The ROC of the GaN lenses is matched to the respective emitter depth and the result for a monolithic diamond hemisphere\,("FIB lens") is shown for comparison.}
\label{fig:PowerandOutlook}
\end{figure}

To quantify this enhancement caused by the SILs, power saturation measurements are taken on two NV$^-$ center pairs before and after SIL printing using the air objective lens with NA\,$=0.95$. The results are shown in Fig.\,\ref{fig:PowerandOutlook}\,a) and as expected, the count rates of the NV centers increase with increasing laser power up to a certain saturation level, after which the count rates plateau. Pair\,1 corresponds to the emitter discussed in Fig.\,\ref{fig:Fluorescence}, while pair\,2 is shown in the bottom right quadrant of the region discussed in Fig.\,\ref{fig:Fabrication}. More information on both emitters and the applied lens profiles can be found in Fig.\,B.6 in the Supplementary Information.
The data is fitted with the sum of two saturation curves derived for two-level quantum systems, allowing both emitters in the focal volume their own saturation power $P_{\mathrm{sat}}$ and intensity $I_{\mathrm{sat}}$:
\begin{equation}
I(P) = \frac{I_{\mathrm{sat1}} \cdot P}{P + P_{\mathrm{sat1}}} + \frac{I_{\mathrm{sat2}} \cdot P}{P + P_{\mathrm{sat2}}}
\end{equation}
A single saturation curve fit describes the data equally well. With this, ($2.2\pm0.1$)$\times$ and ($1.8\pm0.1$)$\times$ enhancement of $I_{\mathrm{sat}}$ is found for pair 1 and pair 2, respectively. Similarly a slight reduction in saturation pump power by a factor of ($1.7\pm0.3$)$\times$ for pair 1 and ($1.2\pm0.2$)$\times$ for pair 2 is observed. This is likely due to reduction in the spherical aberration that occurs at a planar interface with high index contrast\,\cite{Jamali2014}. We do not expect additional substantial narrowing of the point spread function of the pump laser caused by the SIL because objective lens and SIL exhibit similar numerical aperture, compare Fig.\,B.6 in the Supplementary Information. But the pump efficiency enhancement would likely be more noticeable when using a lower NA objective, because the SIL is then expected to reduce the diffraction limited spot size.\\
These measurement results are compared to the expected collection enhancement derived from the simulations shown in Fig.\,\ref{fig:Motivation}. The transmitted light through the detector surfaces (green lines) in the air above the diamond substrate / GaN SIL are averaged between 650 and 750\,nm wavelength. The far field projection is similarly averaged over 650, 700 and 750\,nm wavelength. As neither transmission nor angular dependency vary strongly with wavelength, both can be multiplied to estimate the light collection efficiency in dependence of the acceptance angle / NA of the collection optics, without the need to weight the simulation result with the spectral density of the NV$^-$ emission. The simulation result is shown in Fig.\,\ref{fig:PowerandOutlook}\,b), comparing the planar diamond surface to a GaN micro-lens with its radius of curvature matched to the emitter depth and ideal lateral alignment. The integrated transmission refers to the percentage of total light emitted by the dipole that is caught within the respective angular acceptance cone. The simulations take Purcell enhancement into account, which is found to be $<5\,\%$ within the given spectral region.

The simulations predict an improvement factor of around 2.5$\times$ for an objective lens with NA\,$=0.95$, which is a bit higher than what is found in the experiment. For both NV$^-$ center pairs the real lateral displacement is less than the predicted critical value of $\pm1\,\mu$m, but vertical misalignment might lead to lower collection enhancement, if the emitter is actually placed deeper inside the crystal than intended. We estimate that an emitter that lays 1\,$\mu$m too low, could cause the enhancement to drop to around a factor of 2$\times$, compare Fig.\,B.4 in the Supplementary Information. Note that the SIL placed above NV$^-$ center pair\,1 was chosen to have a slightly larger diameter, leading to a larger ROC, for which the simulations predict increased collection at large collection NA, compare Fig.\,B.5 in the Supplementary Information.\\
As noted before, the simulations for the GaN SIL predict similar overall collection efficiency for objective lenses with NA varying between 0.6 and 0.95, potentially gaining a larger field of view without loss of photon count rate. Taken together, these results provide further evidence that the GaN SILs are effective in enhancing the signal from quantum defects in diamond.\\
Further enhancement could be achieved by placing the color center closer to the diamond surface and increasing the GaN epilayer thickness. In our current work, the 5\,$\mu$m depth of the NV emitters, the 2\,$\mu$m thick AlGaN/AlN buffer layer and the epilayer constrained GaN lens height\,(2\,$\mu$m) limit the enhancement expected from the SILs, because these parameters affect how much the lens aperture covers the angular space above the emitter. The simulated enhancement factors for ROC-matched GaN lenses combined with emitters at various depths and varying GaN lens height compared to a hemispherical monolithic SIL are depicted in Fig.\,\ref{fig:PowerandOutlook}\,c). These results show that a 2\,$\mu$m high GaN micro-lens is expected to increase the light collection efficiency up to 7$\times$\,(NA\,$=0.5$), 6$\times$\,(NA\,$=0.7$) and 4$\times$\,(NA\,$=0.95$), if the emitter is placed in 1\,$\mu$m proximity to the surface, compare with the light gold colored curve in Fig.\,\ref{fig:PowerandOutlook}\,c). In particular the overall absolute collection is not expected to deviate much between an NA of 0.7 and 0.95, potentially offering larger field of view without any losses if a suitable GaN SIL is used. But we note that with moving both to lower NA and lower emitter depth the confocal rejection of the PL from the GaN/AlGaN/AlN layer stack is likely to decrease. In addition, the GaN epilayer thickness could be increased to e.g. 4\,$\mu$m, which allows to increase the lens diameter while maintaining the midpoint position of the spherical lens profile. Thus, the effective angular coverage of the lens aperture above the emitter would rise, improving the photon extraction regardless of emitter depth, compare the dark green and dark golden curves in Fig.\,\ref{fig:PowerandOutlook}\,c) with the light colored curves. This approach is expected to require tuning of the strain profile in the epilayer including the redesign of the buffer layer\,\cite{Wessling2022}.\\
Overall, the derived collection efficiency for the planar diamond surface and the monolithic diamond SIL are in good agreement with previous experimental and theoretical work\,\cite{Hadden2010,marseglia2011nanofabricated,Jamali2014,jiang2014focused,baba1999aberrations,siyushev2010monolithic,Rogers2014,zwiller2002improved}, indicating the validity of the simulations. We added the expected enhancement and absolute collection efficiency for various emitter depths and different GaN epilayer thickness both for (100) and (111) crystal orientation in Fig.\,B.8 in the Supplementary Information, with the overall trends being very similar to what is shown in Fig.\,\ref{fig:PowerandOutlook}\,c).

\section{Conclusion and Outlook}
In summary, we have investigated how additive GaN micro-lenses can be used to enhance the photon collection from diamond color centers, using laser written negatively charged nitrogen vacancy center as an example. To our knowledge, this is the first work that deterministically combines ca. 10\,$\mu$m large semiconductor micro-lenses with color centres in a foreign host crystal, showing the potential of additive high-NA micro-optical components for quantum technology based systems in general. We find evidence of collection improvement on the order of 2$\times$ in good agreement with finite difference time domain simulations. The main advantages that transfer printed GaN solid immersion lenses offer over monolithic diamond hemispheres fabricated with focused ion beam milling are the potentially much faster fabrication speed and their additive nature that eliminates damage to the diamond lattice which could affect the color center properties.\\
Further improvement of photon collection efficiency is expected by placing the color center closer to the diamond-air interface. We find that a dipole emitter in 1\,$\mu$m proximity to the surface might experience 4-7$\times$ collection improvement. Still, the highest possible collection efficiency offered by these additive lenses remains limited in comparison to monolithic hemispheres due to the 2\,$\mu$m thick buffer layer. This might be mitigated by increasing the GaN epilayer thickness to 4\,$\mu$m, potentially achieving lenses with larger diameter but same midpoint of the lens sphere, covering a larger angular space above the emitter. When using an objective lens with NA\,$=0.7$ in this arrangement, the simulations predict that collection is relatively on par with what can be achieved with a hemispherical diamond lens, using the same collection optics.\\
Deterministic laser writing of color centers has been reported with near unity yield\,\cite{Chen2019}, making regular high quality color center arrays possible which could be combined with regularly spaced micro-optical elements for effective collection improvement. The GaN lens fabrication uses highly parallel ICP etching, enabling thousands of device to be fabricated in one wafer run, potentially utilising 6" GaN-on-Si wafer technology. The stamp-based transfer printing process is conducted manually here, but can easily be automated while maintaining $\mu$m-precise placement accuracy using optically visible marker structures\,\cite{mcphillimy2018high,mcphillimy2020automated}. Combined with a multi-stamp head approach\,\cite{ahmed2014active}, a device throughput of $>100-200$ devices per hour is conceivable after initial alignment of donor and receiver chip and highly optimised processing. Alternatively continuous roller transfer printing might offer a way to scale the transfer process, but still needs further development in terms of overlay alignment accuracy\,\cite{margariti2023continuous}. Further scaling could be achieved by adding multiple GaN micro-lenses on one AlGaN/AlN membrane to create printable arrays, thus reducing the number of transfer print processes. Lens arrays could either be integrated with arrays of deterministically generated color centers or deterministic writing could be performed through the GaN lenses themselves after printing\,\cite{Chen2019}. The latter approach could allow the reduction of depth at which vacancies can be created by laser writing and would additionally auto-align the respective color center close to the midpoint of the lens sphere\,\cite{yurgens2021low}.

\appendix
\section{Backmatter}
\subsection{Funding}
The authors acknowledge funding from the following sources: Royal Academy of Engineering (Research Chairs and Senior Research
Fellowships); Engineering and Physical Sciences Research Council (EP/R03480X/1, EP/N017927/1, EP/P00945X/1, R004803/1, EP/M013243/1, EP/T001062/1, EP/V056778/1, EP/L015315/1); Innovate UK\,(50414);
Fraunhofer Lighthouse Project QMag. NKW acknowledges funding of his PhD studentship by Fraunhofer UK. G. W. M is supported by the Royal Society.\\

\subsection{Acknowledgement}
The authors acknowledge Benoit Guilhabert (University of Strathclyde) for his development work on both the transfer print technique and suspension of GaN-on-Si thin films as well as Luke Johnson (University of Warwick) for acid cleaning the diamond sample after annealing.
\subsection{Disclosures}
The authors declare no conflicts of interest.
\section{Supplementary Information}
This document provides further details on the experimental setups, an analysis of displacement and different lens geometries with FDTD simulations, additional AFM and PL data on the discussed emitters and PL spectra taken on GaN/AlGaN/AlN thin films at room temperature.
\counterwithin{figure}{section}
\setcounter{figure}{0}
\subsection{Confocal setups used for photoluminescence measurements}

\begin{figure}[H]
\centering
\includegraphics[width=\linewidth]{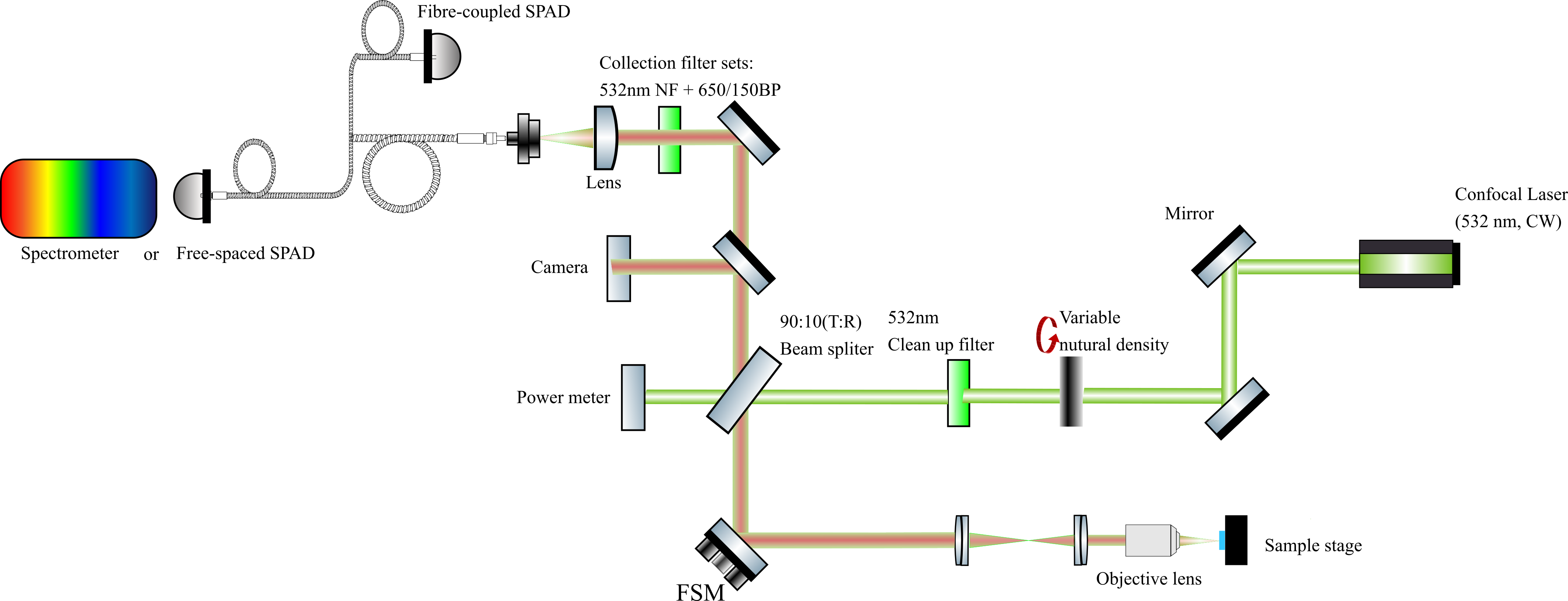}
\caption{Home-built confocal microscope for dual air (NA = 0.95) and oil (NA = 1.25) objective lens}
\label{fig:WASPS setup}
\end{figure}

\begin{figure}[H]
\centering
\includegraphics[width=0.6\linewidth]{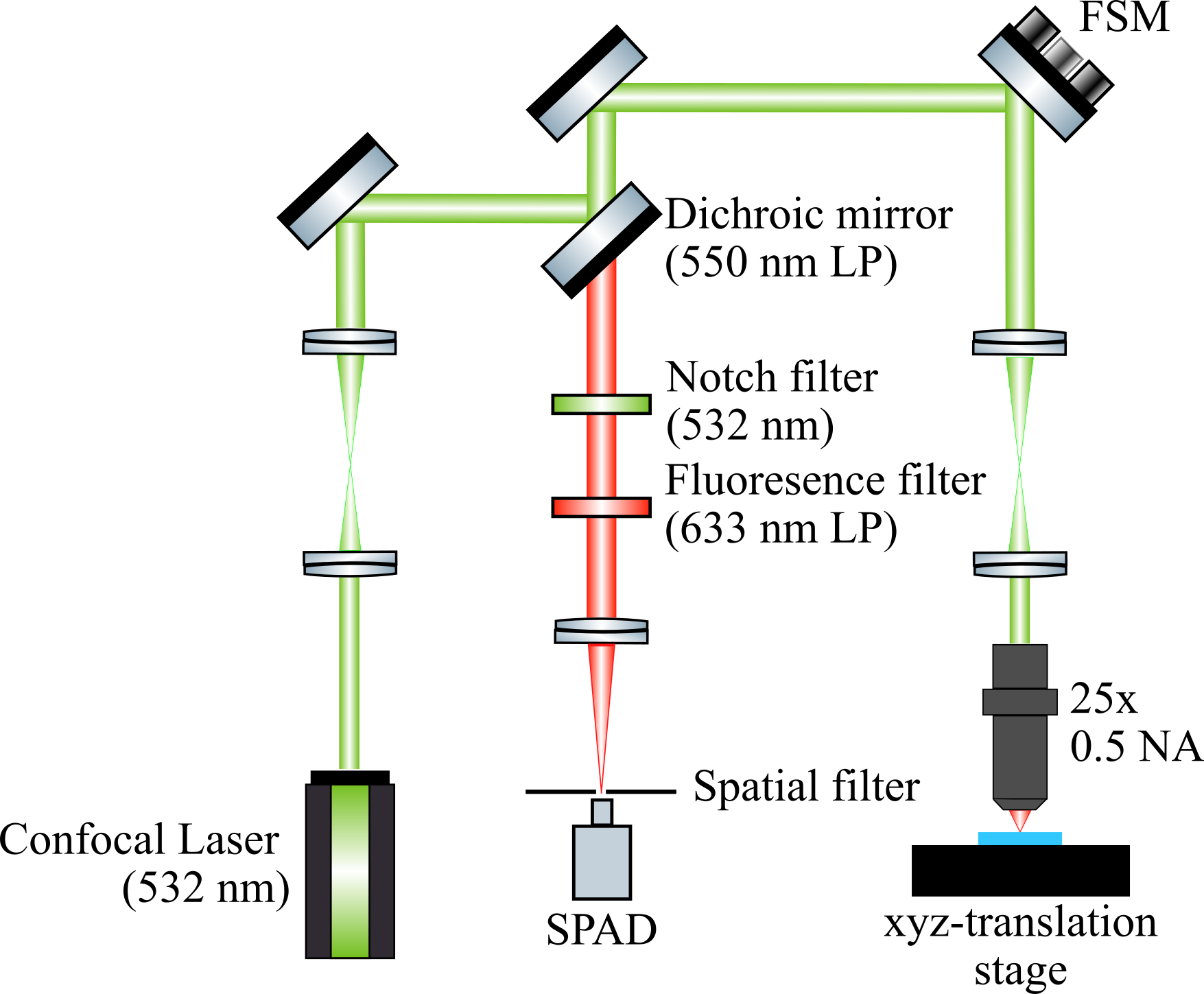}
\caption{Home-built confocal microscope for NA = 0.5}
\label{fig:ENG setup 2}
\end{figure}

The confocal photoluminescence setup that was used to gather the power series on the two doublet emitters in addition to the presented autocorrelation and spectral measurements with a NA = 0.95 air lens is depicted in Fig.\,\ref{fig:WASPS setup}. NA = 1.25 oil lens were used in the same setup to get spatial distribution information of the emitter of interest with best signal to noise ratio. Here, a continous wave 532 nm laser (GEM532) excitation source with 1\,mW output power was used, the fluorescence signal was collected back through the same objectives lens and spectral filtering of the excitation laser is applied before detection. 
\\ Fig.\,\ref{fig:ENG setup 2} shows the setup used for measurements with the NA = 0.5 air objective. A continuous-wave 532\,nm laser (GEM532) with 6\,mW output power was used as excitation source.
Photoluminescence maps are taken with both setups.

\subsection{Displacement of emitter with respect to the mid point of the lens sphere and the influence of an air gap on the collection efficiency}
\begin{figure}[H]
\centering
\includegraphics[width=\linewidth]{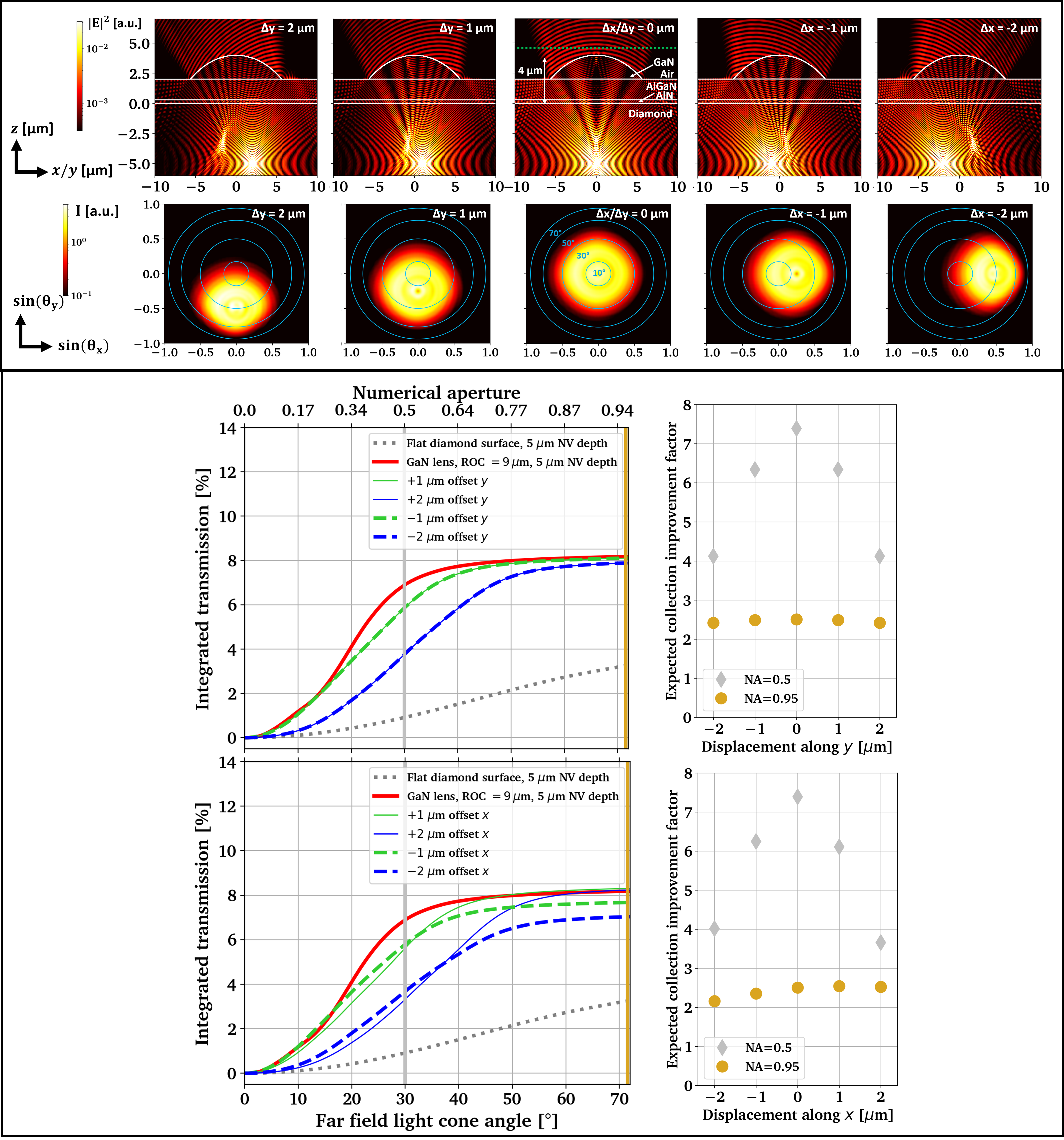}
\caption{Influence of $x$ and $y$ displacement on the expected collection efficiency from the two dipole emitter. The radius of curvature of the GaN micro-lens is matched to the emitter depth of 5\,$\mu$m. Cross sections are taken at $\lambda=700$\,nm wavelength and the transmission is averaged between $\lambda=650-750$\,nm wavelength.}
\label{fig:xyDisplacement}
\end{figure}

\begin{figure}[H]
\centering
\includegraphics[width=\linewidth]{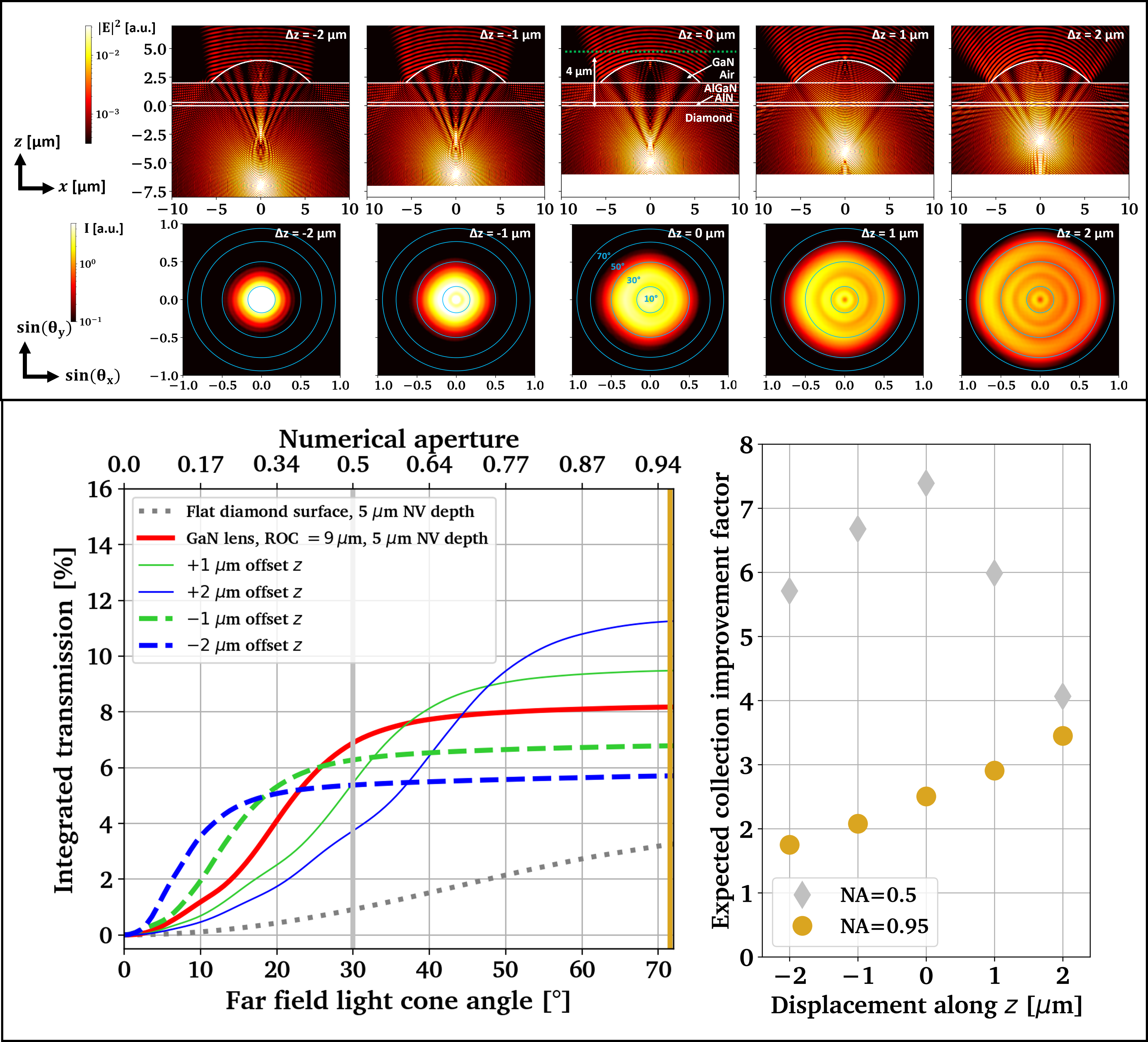}
\caption{Influence of $z$ displacement on the expected collection efficiency from the two dipole emitter. The radius of curvature of the GaN micro-lens is matched to the emitter depth of 5\,$\mu$m. Cross sections are taken at $\lambda=700$\,nm wavelength and the transmission is averaged between $\lambda=650-750$\,nm wavelength.}
\label{fig:zDisplacement}
\end{figure}

\begin{figure}[H]
\centering
\includegraphics[width=0.95\linewidth]{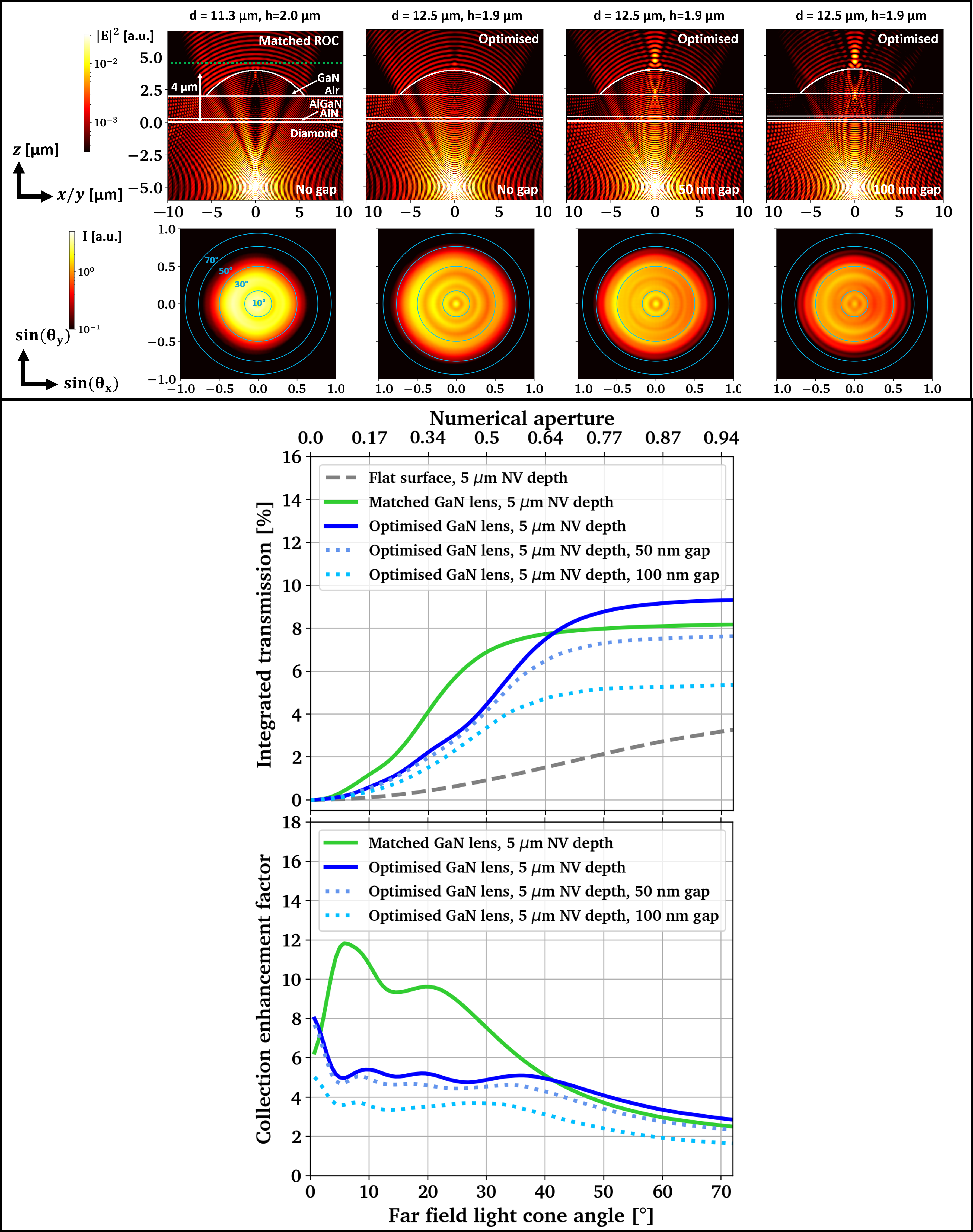}
\caption{FDTD simulations taking the measured AFM profile from the lens above emitter pair 1\,(Fig.\,\ref{fig:AdditionalData},a and Fig.\,4 in the main text) into account. The most left cross section and far field plot show the previously discussed GaN lens with matched radius of curvature for reference. Additionally, the effect of an air gap between diamond and AlN bottom layer of the GaN lens layer stack is investigated on the right hand side of the figure. Cross sections are taken at $\lambda=700$\,nm wavelength and the transmission is averaged between $\lambda=650-750$\,nm wavelength.}
\label{fig:RealShapeandEffectofaGap}
\end{figure}

In this section we present additional FDTD simulations to investigate the effects of small displacements of two dipole emitters mimicking a NV$^-$ centre in a diamond crystal with (100) surface orientation below a GaN SIL which radius of curvature is matched to the emitter depth of 5\,$\mu$m. Two separate simulations are run and averaged to account for the two different dipole moments of the NV$^-$ center, one of which is tilted by (90-54.7)\,$^{\circ}$ with respect to the surface normal\,(the NV axis is tilted by 54.7\,$^{\circ}$, but both dipole moments are orientated perpendicular to the NV axis). Cross sections and far field projection are taken at $\lambda=700$\,nm wavelength. The expected collection efficiency is calculated by averaging the transmission through the detector surface above the lens\,(indicated in green) multiplied with the projected far field distribution from the same detector, choosing $\lambda=650-750$\,nm wavelength to match the spectral emission region of the NV$^-$ centre. Both transmission and far field projection are found to be insensitive towards wavelength changes in this spectral regime, allowing us to ignore the spectral density distribution of the emission spectrum.\\
Lateral displacement is discussed in Fig.\,\ref{fig:xyDisplacement}, vertical displacement in Fig.\,\ref{fig:zDisplacement}. We illustrate the effects of a slightly larger radius of curvature accompanied by a larger diameter of the micro-lens on the collection improvement as function of an air gap between diamond and the AlN bottom surface of the lens platelet in Fig.\,\ref{fig:RealShapeandEffectofaGap}.

\subsection{Additional photoluminescence and AFM measurements of SILs and emitters}
\begin{figure}[H]
\centering
\includegraphics[width=\linewidth]{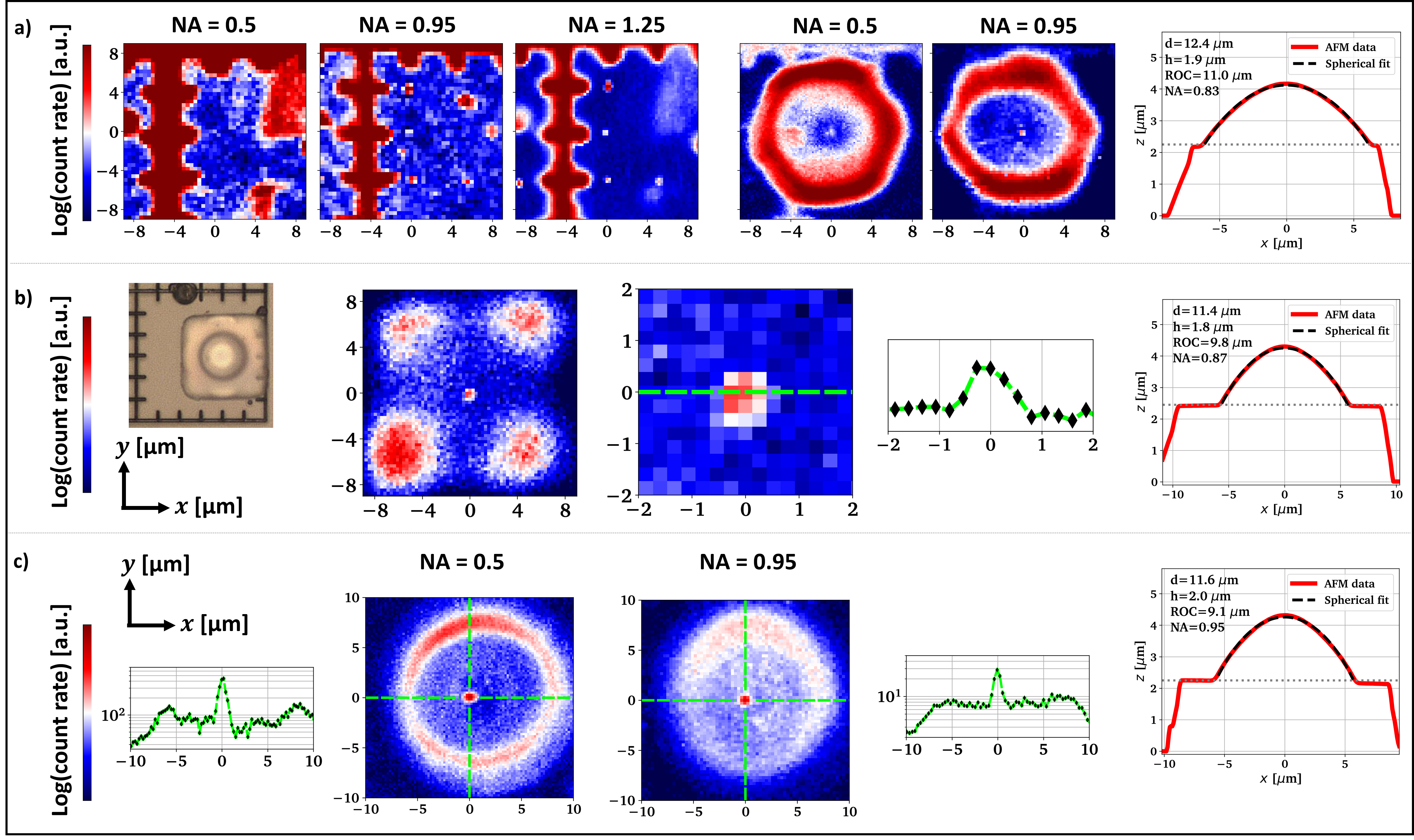}
\caption{Photoluminescence and AFM measurements regarding a) doublet emitter 1 shown in Fig.\,4, b) doublet emitter 2 included in the power saturation measurement shown in Fig.\,5\,a), c) the micro-lens in the top left quadrant discussed in Fig.\,3.}
\label{fig:AdditionalData}
\end{figure}

\begin{figure}[H]
\centering
\includegraphics[width=\linewidth]{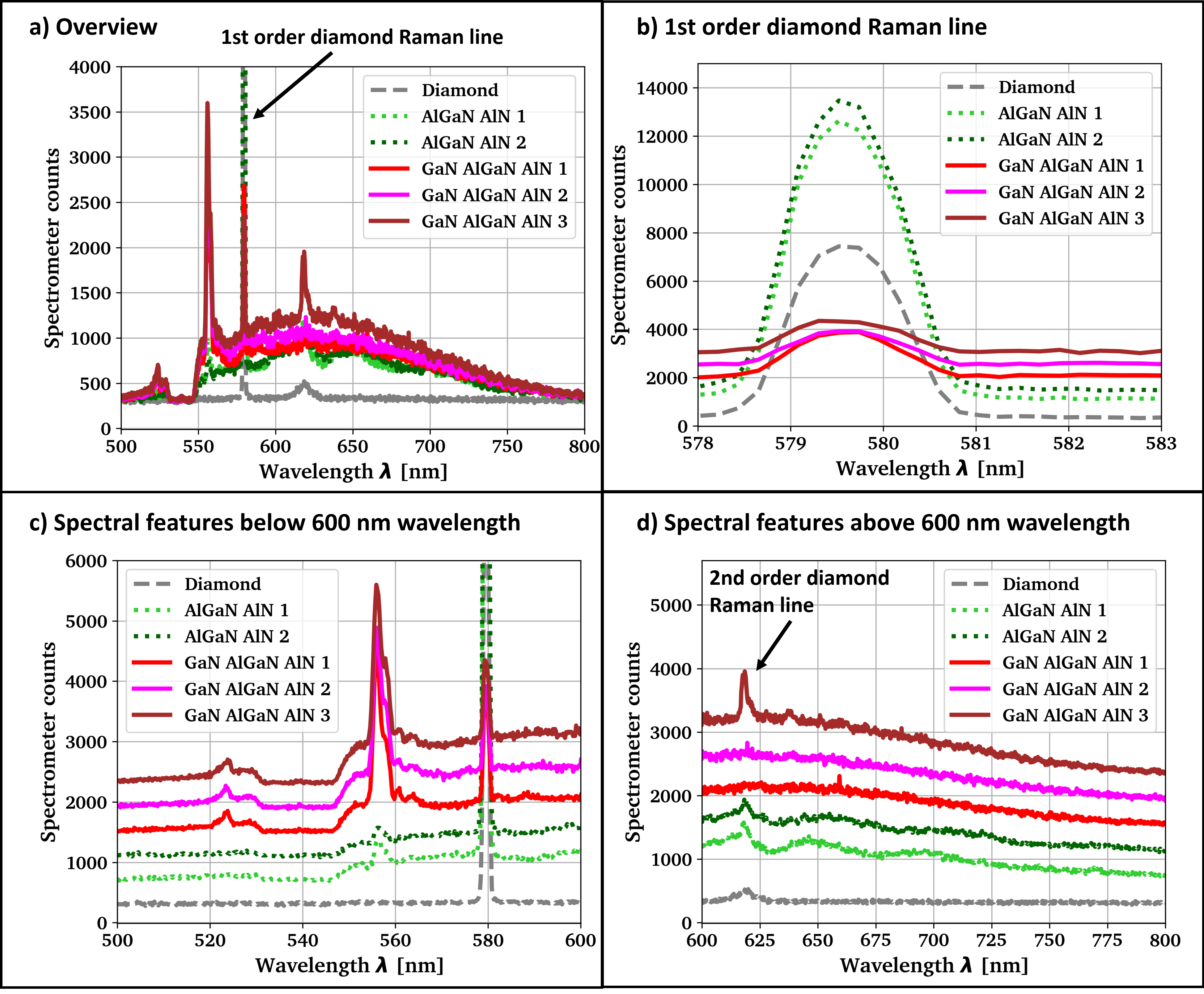}
\caption{Room temperature photoluminescence spectra of transferprinted GaN lenses on AlGaN/AlN mesa structures on a single crystalline CVD grown polished diamond sample. a)-d) show the same spectra, which are separated linearly in b)-d) for better visibility.}
\label{fig:AdditionalData_PLSpectrum}
\end{figure}

To provide some additional data from the investigated emitters, Fig.\,\ref{fig:AdditionalData}\,a) includes a larger field of view of the photo luminescence plots shown in Fig.\,4\,a) in the main text as well as an AFM line scan of this slightly larger SIL. Fig.\,\ref{fig:AdditionalData}\,b) includes photoluminescence maps and an AFM line scan of the second doublet emitter discussed in less detail in the main text, while Fig.\,\ref{fig:AdditionalData}\,c) illustrates the NA dependency of the PL map taken on the graphitised emitter spot discussed in Fig.\,3 in the main text.\\
Room temperature spectral measurements of transferprinted GaN lenses and AlGaN/AlN membranes on single crystalline diamond are shown in Fig.\,\ref{fig:AdditionalData_PLSpectrum}. A 532\,nm CW excitation laser is used with a 550\,nm long pass filter and 532\,nm notch filter in a confocal arrangement. 

\subsection{Detailed FDTD simulation results for emitters at different depth and increased GaN epilayer thickness}
\begin{figure}[H]
\centering
\includegraphics[width=0.98\linewidth]{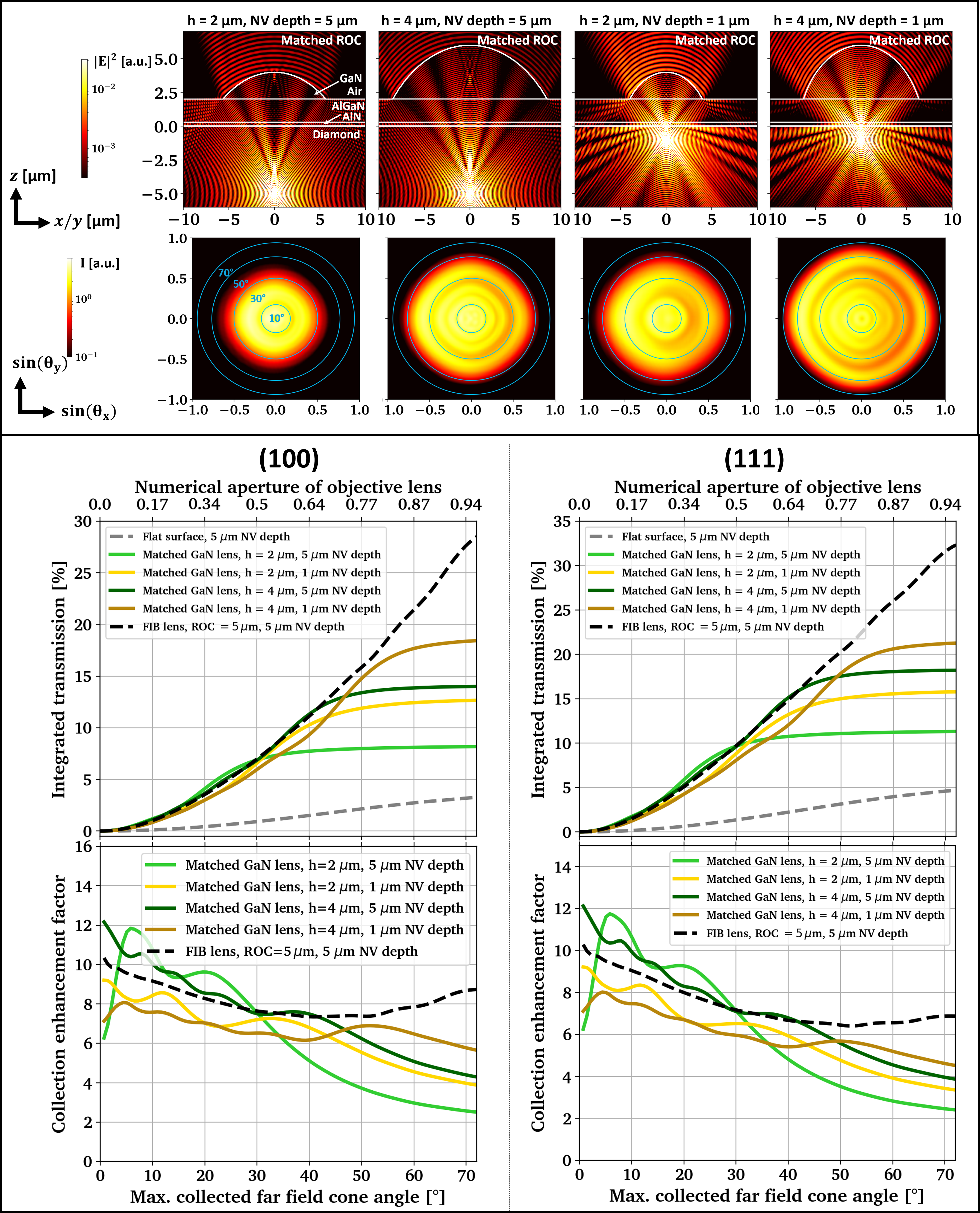}
\caption{Light collection and its enhancement in dependence of emitter depth and GaN lens height expected form FDTD simulations. The emitter position is matched to the geometric center of each lens. The cross sections are taken for (100) diamond surface orientation, while the bottom plots distinguish between (100) and (111), assuming an ideally aligned emitter in the (111) case. Cross sections are taken at $\lambda=700$\,nm wavelength and the transmission is averaged between $\lambda=650-750$\,nm wavelength.}
\label{fig:Enahncement}
\end{figure}

To provide a more detailed overview of the simulation results for dipole emitters in different depth below the diamond surface and the potential arising from a thicker GaN epilayer, we summarised the absolute collection efficiency and collection enhancement as function of the NA of the collection optics for both (100) and (111) crystal direction in Fig.\,\ref{fig:Enahncement}. The bottom left plot is already shown in the main text in Fig.\,5\,c).

\newpage
\bibliographystyle{ieeetr}
\bibliography{ms}  

\end{document}